\documentclass[12pt]{article}

\usepackage{epsfig}

\def\bomega{\mbox{\boldmath $\omega$}}
\def\balpha{\mbox{\boldmath $\alpha$}}
\def\bbeta{\mbox{\boldmath $\beta$}}
\def\btau{\mbox{\boldmath $\tau$}}
\def\bnu{\mbox{\boldmath $\nu$}}

\def\bpsi{\mbox{\boldmath $\psi$}}

\def\bdelta{\mbox{\boldmath $\delta$}}

\def\bV{\mbox{\bf V}}
\def\bZ{\mbox{\bf Z}}

\def\bPhi{\mbox{\boldmath $\Phi$}}

 \def\tfrac#1#2{{\textstyle{{#1}\over{#2}}}}

\def\half{\tfrac{1}{2}}
\def\third{\tfrac{1}{3}}
\def\quarter{\tfrac{1}{4}}
\def\sixth{\tfrac{1}{6}}
\def\twothirds{\tfrac{2}{3}}
\def\threequarters{\tfrac{3}{4}}

\def\Tr{\mathop{\rm Tr}\nolimits}

\usepackage{amssymb}

\begin{document}

\begin{titlepage}

\baselineskip 24pt

\begin{center}

{\Large {\bf Unified FSM treatment of CP physics extended 
to hidden sector giving (i) $\delta_{CP}$ for leptons as 
prediction, (ii) new hints on the material content of the 
universe}}

\vspace{.5cm}

\baselineskip 14pt

{\large Jos\'e BORDES \footnote{Work supported in part by Spanish
    MINECO under grant FPA2017-84543-P and PROMETEO 2019-113 (Generalitat Valenciana).}}\\
jose.m.bordes\,@\,uv.es \\
{\it Departament Fisica Teorica and IFIC, Centro Mixto CSIC, Universitat de 
Valencia, Calle Dr. Moliner 50, E-46100 Burjassot (Valencia), 
Spain}\\
\vspace{.2cm}
{\large CHAN Hong-Mo}\\
hong-mo.chan\,@\,stfc.ac.uk \\
{\it Rutherford Appleton Laboratory,\\
  Chilton, Didcot, Oxon, OX11 0QX, United Kingdom}\\
\vspace{.2cm}
{\large TSOU Sheung Tsun}\\
tsou\,@\,maths.ox.ac.uk\\
{\it Mathematical Institute, University of Oxford,\\
Radcliffe Observatory Quarter, Woodstock Road, \\
Oxford, OX2 6GG, United Kingdom}

\end{center}

\vspace{.3cm}

\begin{abstract}

A unified treatment of CP physics for quarks and leptons in the framed
standard model (FSM) is extended to include the predicted hidden sector 
giving as consequences: 
(i) that an earlier part-estimate of the Jarlskog invariant $J'$ for 
leptons is turned into a prediction for its actual value, i.e.,
$J' \sim -0.012$ ($\delta'_{CP} \sim 1.11 \pi$), which is of the right 
order of magnitude, of the right sign, and in the range of values 
favoured by present experiment,  
(ii) some novel twists to the effects of CP-violation on the material 
content of the universe.

\end{abstract} 

\end{titlepage}

\newpage

\section{Preamble}

\subsection{CP in the FSM}

The framed standard model (FSM) \cite{fsmpop,tfsm}, constructed 
initially for understanding the mass and mixing patterns of quarks 
and leptons, is found to provide also a neat unified treatment of 
CP physics in the world of quarks and leptons covered by the 
standard model (SM).  

Briefly, this has come about as follows. The mass matrix for quarks 
and leptons in FSM turns out to be of the following form:
\begin{equation}
m = m_T \balpha \balpha^\dagger,
\label{mfact}
\end{equation}
where the scalar $m_T$ depends on fermion type, but the real unit 
vector $\balpha$ in 3-dimensional generation space is the same for 
all quarks and leptons, and changes with scale (rotates) in a manner 
governed by renormalization group equations (RGE).  And it is the 
rotation of this $\balpha$, according to the FSM, which gives, and 
quite accurately, the intricate mass and mixing pattern for quarks 
and leptons which is observed in experiment.

As for CP and its violations, the usual SM treatment has some loose 
ends:
\begin{itemize}
\item {\bf [a]} What is known as the strong CP problem, namely in 
the QCD action a theta-angle term of topological origin (instantons) 
\cite{thooftinst,Weinbergqft} which potentially can give CP-violating 
effects many orders of magnitude larger than is allowed by experiment 
\cite{edmneutron}.
\item {\bf [b]} A CP-violating Kobayashi-Maskawa (KM) phase \cite{KM} 
is formally allowed in the CKM mixing matrix for quarks but the size 
and physical origin of which are left unexplained.
\item {\bf [c]} Although a theta-angle term similar to {\bf[a]} is in 
principle present also in the flavour action, there are suggestions 
that it can be trivially transformed away\footnote{based on arguments 
we find insufficient, see \cite{cpslept})}.
\item {\bf [d]} A CP-violating phase, similar to {\bf [b]}, can exist 
also in the PMNS mixing matrix for leptons, but again with size and 
origin equally unexplained.
\end{itemize}

However, the form of the mass matrix (\ref{mfact}) in the FSM changes 
all that.
\begin{itemize}
\item {\bf [a$'$]} The mass matrix (\ref{mfact}) has a state in the 
direction of the Darboux binormal $\bnu$ with zero mass eigenvalue 
\footnote{When we consider a curve embedded on a given surface, as in 
the case of the trajectory of $\balpha$ on the unit sphere, the 
radial vector $\balpha$, the tangent $\btau$ to the curve, and the 
binormal $\bnu$ form an orthonormal triad which we call the Darboux 
triad.  Note that the choice above of the zero eigenvalue direction 
along the $\bnu$ direction is unique.  Although (\ref{mfact}) has a 
zero eigenvalue also in the direction of $\btau$, a chiral 
transformation in that direction at any $\mu$ would make the vector 
$\balpha(\mu + \delta \mu) = \balpha(\mu) + \delta \mu \btau$ 
complex at a neighbouring scale while the RGE governing its rotation 
would keep $\balpha$ real \cite{strongCP,atof2cps}.}.  It follows 
then that the theta-angle term can be cancelled in Feynman path 
integrals by a chiral transformation on that zero mode, solving thus 
the strong CP problem at every scale $\mu$.
\item {\bf [b$'$]} But, since $\balpha$ rotates with scale, the 
direction of $\bnu$ also changes with scale; the cancellation in 
{\bf [a$'$]} at one scale does not guarantee the same at another 
scale.  This scale-dependence figures in the CKM matrix (which 
compares the direction of up and down state vectors measured at 
different scales) and appears as the KM phase, providing thus not 
only an explanation of its origin but also a means to calculate its 
size \cite{strongCP,atof2cps,tfsm}.
\item {\bf [c$'$]} Parallel to the considerations {\bf [a$'$]} for 
quarks but now applied instead to leptons cancels the effect of the 
theta-angle term in flavour.
\item {\bf [d$'$]} By the same token as {\bf [b$'$]} the scale 
dependence of $\balpha$ gives a CP-violating phase in the PMNS matrix 
for leptons, explains its origin and gives its size.
\end{itemize}
Besides it has been shown \cite{cpslept} that the CP-violations as 
measured by the Jarlskog invariants $J$ \cite{jarlskog} (equivalent 
in content to the phases $\delta_{\rm CP}$ but easier to work with 
in FSM because parametrization invariant) so obtained in {\bf [b$'$]} 
and {\bf [d$'$]} are both of the same order of magnitude as those 
observed in experiment so that no other sources for CP-violation are 
needed for their explanation at present. 

The result is, it seems,
\begin{itemize}
\item {\bf [CPU]}  A unified treatment of all at present known 
effects of CP-violation in particle physics, in which:
\begin{itemize}
\item CP-violations for quarks and leptons are put on the same 
footing.
\item They both have a topological origin in the theta-angle 
terms in respectively the colour and flavour theory.
\item In both cases, potentially huge violations in theta-angle 
terms are converted to CP-violating phases in the mixing matrices 
of the order seen in experiment.
\end{itemize} 
\end{itemize}

If true, this is considerably neater than what we had for the SM in 
{\bf [a]}---{\bf [d]} above, thus tidying up the loose ends that 
have been left untied.

\subsection{The hidden sector}

However, neater as this might seem, it is for the FSM still incomplete 
because, in addition to the particles already known to us, designated 
here as the standard sector {\bf [SS]}, the FSM predicts another, what 
we might call 
\begin{itemize}
\item {\bf [HS]} the hidden sector, which communicates but little with 
the standard sector we know, and turns out to be related to the latter 
by having the roles of colour and flavour interchanged \cite{fsmpop,cfsm}, 
as will be explained immediately below. 
\end{itemize}
Given further that the stable neutral members of this hidden sector are 
all potential candidates for dark matter, it is clear that {\bf [HS]}, 
if true, would have far-reaching consequences, and any consideration in 
FSM of CP or its violation would be incomplete without taking account 
of it.
  
That a hidden sector {\bf [HS]} should exist in FSM comes about as 
follows.  The FSM is constructed by framing the standard model, meaning 
that the frame vectors (vielbeins) in the internal symmetry space are 
promoted into fields and treated as dynamical variables alongside the 
usual gauge boson and matter fermion fields.  These new fields, called 
framons, though vectors in internal symmetry space, are Lorentz scalars.  
Specifically, they appear as:
\begin{itemize}
\item {\bf [FF]} The flavour (``weak'') framon:
\begin{equation}
\balpha \Phi = \alpha^{\tilde{a}} \phi_r^{\tilde{r}},
\label{PhiF}
\end{equation}
\item {\bf [CF]} The colour (``strong'') framon:
\begin{equation}
\bbeta \bPhi = \beta^{\tilde{r}} \phi_a^{\tilde{a}},
\label{PhiC}
\end{equation}
\end{itemize}
where:
\begin{equation}
r = 1, 2, \ \ \tilde{r} = \tilde{1}, \tilde{2}, \ \ 
a = 1, 2, 3, \ \ \tilde{a} =\tilde{1}, \tilde{2}, \tilde{3}.
\end{equation}
The $2 \times 2$ matrix $\Phi$ and the $3 \times 3$ matrix $\bPhi$ are 
scalar fields depending on space-time points $x$, while the 3 vector 
$\balpha$ and 2 vector $\bbeta$ are $x$-independent global quantities 
\cite{efgt}.  

The flavour framon, of which only one column needs to be kept, as 
explained in for example \cite{tfsm}, is, apart from the factor 
$\balpha$, essentially just the Higgs field of the standard electroweak 
theory.  But, the colour framon is entirely new with no analogue in the 
standard model.  And the hidden sector we are after is basically just 
the manifestation of the new degrees of freedom that the colour framons 
{\bf [CF]} represent.

These colour framons are colour triplets, and since colour is confined, 
they cannot propagate freely as particles in space, but have to combine 
with other coloured objects to form colour singlets.  A colour framon 
can combine with its own conjugate to form scalars (generically denoted 
by H) or vectors (generically denoted by G), or with colour triplet 
fermions to form colour neutral fermion states (generically denoted by F) 
\cite{cfsm}.  They are the analogues of respectively the Higgs boson 
$h_W$, the vector bosons $\gamma$-$Z$, $W$, and the quarks and leptons 
in the flavour theory when the latter is pictured as a confining theory, 
as suggested by 't~Hooft \cite{tHooftcon}.  So, as our standard sector 
is populated by $h_W$, $\gamma$-$Z$, $W$ and quarks and leptons, the 
hidden sector is populated by the H, G, and F.  Some details known so 
far of these predicted particles H, G and F are spelt out in \cite{cfsm}, 
and possible evidences for the existence of some of them have been 
examined in \cite{zmixed,fsmanom}.  

Given then that this hidden sector exists in the FSM, our consideration 
of CP and its violations as summarized in {\bf [CPU]} would be incomplete 
unless extended to include the hidden sector as well.

\section{Yukawa couplings for colour framons}

An extension into the hidden sector of the analysis of CP and its violation,
along the lines one did for the standard sector \cite{cpslept} as summarized 
in {\bf [CPU]} above, is not going to be straightforward.  The arguments for 
that analysis, one recalls, centre on the fermion mass matrix, that is, on 
the Yukawa coupling, but the Yukawa coupling for the hidden sector is 
unfortunately the weakest link in the theoretical structure of the FSM, as 
emphasized in \cite{cfsm}.

The reason is the following.  Of the 3 sets of fields functioning as dynamical 
variables of the FSM, the matter fermion fields are the ones of which we know 
the least.  The vector bosons are gauge fields, or connections of fibre bundles, 
with a clear geometrical meaning and a particular function to discharge.  What 
they are, and how they should enter in the FSM action harbour therefore little 
doubt.  The framon scalars also, once ascribed the geometrical significance of 
frame vectors in internal symmetry space, possess likewise a geometrical 
meaning and function, and pose again few questions on what they should be and 
how they should enter in the FSM action.  For the matter fermion fields, on the 
other hand, one does not know as yet of the geometrical role, if any, they play.  
One does not know {\it a priori} even what matter fermion fields should enter in 
the theory, let alone how they are coupled to one another.  This diffculty was 
there, of course, already in the standard sector, but there one had the benefit 
of experimental knowledge, by dint of which, and of the ingenuity and patience 
of pioneer workers, one knows what fermion fields should enter in the theory, 
namely 3 generations each of flavour doublet left-handed and flavour singlet 
right-handed quarks, and the same for leptons, and hence how to construct the 
Yukawa coupling from them.  But now in the hidden sector, there is of course 
no experiment to help.  One's sole guidance, apart from general invariance 
principles, is analogy with the standard sector.  And with these alone in the 
hidden sector to explore CP and its violations, one is obviously treading on 
rather thin ice.  Any findings achieved thereby can thus be taken as only 
tentative and are to be treated with caution.  It is with this understanding 
that we proceed in what follows.

Now in \cite{cfsm}, a working model was suggested for the Yukawa couplings in 
the hidden sector, that is, the couplings of the colour framon {\bf [CF]} 
to the following tentatively proposed list of fundamental fermion fields:
\begin{eqnarray}
\psi_L(\sixth, 2, 3), \psi_R(-\third, 1, 3), \psi_R(\twothirds, 1, 3),
\nonumber \\
\psi_L(-\half, 2, 1), \psi_R(-1, 1, 1), \psi_R(0, 1, 1) \nonumber \\
\psi_R(-\half, 2, 1), \psi_L(-1, 1, 1), \psi_L(0, 1, 1),
\label{fundferm}
\end{eqnarray}
where the first argument inside the brackets denotes the $u(1)$ charge, the 
second the dimension of the $su(2)$ representation and the third that of the 
$su(3)$ representation.  Of these, those in the first row carry colour and 
can combine with the colour framon via colour confinement to form compound 
Fs of the handedness indicated.  From the first, $\psi_L(\sixth, 2, 3)$, 
having neutralized its colour by the framon, one obtains a bound state F, 
say $\chi_L(\pm \half, 2, 1)$, carrying still local flavour, which we call 
a co-quark, this being the parallel of a left-handed quark in the standard 
sector under colour-flavour interchange~\footnote{The left-handed quark, 
we recall, is in 't~Hooft's confinement picture of the electroweak theory 
\cite{tHooftcon}, a bound state via flavour confinement of a fundamental 
fermion with a Higgs scalar---here a flavour framon---which carries still 
local colour.}.  Similarly, from the other two coloured fields on the list 
(\ref{fundferm}), one obtains two Fs, neither of which carries local 
flavour, and both, being parallels of the right-handed lepton in the 
standard sector, are called co-leptons.    

The left-handed co-quark $\chi_L(\pm \half, 2, 1)$ can then be partnered by
two other members on the list (\ref{fundferm}), namely $\psi_R(-\half, 2, 1)$ 
and $\psi_L(-\half, 2, 1)^C$ (where superscript $C$ denotes charge conjugate) 
to form Yukawa couplings for co-quarks in analogy to those for quarks in the 
standard sector.  Explicitly, in \cite{cfsm}, equation (73), the following 
(with minor changes in notation) were suggested:
\begin{equation}
{\cal A}_{YQ} = \left[ \begin{array}{lll} 
Z_1 \bar{\psi}_L(\sixth, 2, 3) (\bPhi \cdot \bdelta_1) \psi_{R1}(\half, 2, 1) \\
+Z_2 \bar{\psi}_L(\sixth, 2, 3) (\bPhi \cdot \bdelta_2) \psi_{R2}(\half, 2, 1) \\
+Z_3 \bar{\psi}_L(\sixth, 2, 3) (\bPhi \cdot \balpha) \psi_R(-\half, 2, 1) 
\end{array} \right] + {\rm h.c.}
\label{calAYQold}
\end{equation}

Of these, the term proportional to $Z_3$ seems unambiguous, but for the 
other two, some questions remain, which we already noted at the time we 
wrote them down but had not then the incentive to sort out.  Now however, 
with new incentive and hindsight, two shortcomings to these terms can be 
identified:
\begin{itemize}
\item [S1] Before these couplings (\ref{calAYQold}) were inserted, the 
FSM vacuum was invariant under, say, an $\widetilde{su}(2)_H$ symmetry 
in the two directions orthogonal to $\balpha$ \cite{tfsm}.  This is 
broken by (\ref{calAYQold}) explicitly along two arbitrarily chosen 
directions $\bdelta_1$ and $\bdelta_2$.
\item [S2] The two right-handed fields, $\psi_{R1}(\half, 2, 1)$ and 
$\psi_{R2}(\half, 2, 1)$ could, as far as quantum numbers are concerned, 
be coupled to both the $\bdelta_1$ and $\bdelta_2$ terms, but had been 
taken arbitrarily in (\ref{calAYQold}) to couple respectively to only 
one each of these two terms.
\end{itemize}
 
In view of these shortcomings in (\ref{calAYQold}), we propose now to 
replace it by:
\begin{equation}
{\cal A}_{YQ} = \left[ \begin{array}{ll} 
\bar{\psi}_L(\sixth, 2, 3) (\bPhi \cdot \bomega) 
   (\bZ \cdot \bpsi_R(\half, 2, 1)) \\
+ Z_3 \bar{\psi}_L(\sixth, 2, 3) (\bPhi \cdot \balpha) \psi_R(-\half, 2, 1)
\end{array} \right] + {\rm h.c.},
\label{calAYQnew}
\end{equation}
where $\bomega$ is a unit 3-vector in dual colour space (say, orthogonal 
to $\balpha$ and therefore a linear combination of $\bdelta_1, \bdelta_2$), 
while
\begin{equation}
(\bZ \cdot \bpsi_R(\half, 2, 1))  = 
   Z_1 \psi_{R1}(\half, 2, 1) + Z_2 \psi_{R2}(\half, 2, 1).
\label{bdelta+}
\end{equation}
Since both $\psi_{R1}$ and $\psi_{R2}$ can now in general be coupled to 
both $\bdelta_1$ and $\bdelta_2$, the new coupling (\ref{calAYQnew}) can 
claim to have avoided the shortcoming [S2] of the old (\ref{calAYQold}).  
Part of [S1] seems to remain in that the $\widetilde{su}(2)_H$ symmetry 
is still explicitly broken, now along the direction $\bomega$, but if the 
latter rotates (changes in direction) with changing scale, as we shall 
argue below, then at least some democracy is restored between the various 
directions covered by $\widetilde{su}(2)_H$, and [S1] therefore, in this 
aspect, alleviated.  Besides, if $\bomega$ can be assigned a physical 
significance, on which we have some idea not as yet substantiated, then 
the direction in which the $\widetilde{su}(2)_H$ symmetry is broken is 
no longer arbitrary, as were $\bdelta_1$ and $\bdelta_2$ in [S1]. 

Whether these improvements of the new Yukawa coupling (\ref{calAYQnew}) 
on the old version (\ref{calAYQold}) are sufficient to qualify it for 
being taken now as {\it the} Yukawa coupling for the co-quarks is not 
easy to ascertain, all our arguments being based just on analogy with 
the Yukawa couplings of the flavour framon in the standard sector, but 
these latter offered no parallel of the situation here of two $\chi_L$ 
states with identical quantum numbers.  

However, we find (\ref{calAYQnew}) attractive because, as we shall show
below:
\begin{itemize}
\item {\bf [P1]} it gives a mass matrix for co-quarks with a zero mode,
\item {\bf [P2]} the direction of that zero mode might depend on scale,
\end{itemize}
which, as outlined in Section 1.1. and detailed in \cite{cpslept}, were 
the conditions which cured the strong CP problem in the standard sector, 
and which we hope may do the same for the parallel problem that will
later appear in the hidden sector as well.  We accept (\ref{calAYQnew}) 
therefore as now our working model. 

That {\bf [P1]} attains can be seen as follows.  The Yukawas coupling 
(\ref{calAYQnew}) gives for the 2 charged $\half$ co-quarks a factorizable 
mass sub-matrix proportional to:
\begin{equation}
\hat{m} \sim \left( \begin{array}{c} \omega_1 \\ \omega_2 \end{array} \right)
             (Z_1, Z_2) \half(1 + \gamma_5) + {\rm h.c.},
\label{mhat}
\end{equation}
where $(\omega_1, \omega_2)$ are the components of $\bomega$ at any chosen 
scale along the two directions $\bdelta_1$ and $\bdelta_2$.  This submatrix 
has a zero mode in the direction orthogonal to $(\omega_1, \omega_2)$, as 
can be seen as follows.  By a suitable relabelling of the right-handed 
fields, following a procedure made familiar already with the mass matrix 
$m$ in (\ref{mfact}) for quarks and leptons, $\hat{m}$ of (\ref{mhat})
can be cast into a Hermitian form independent of $\gamma_5$, thus:
\begin{equation}
\hat{m} \rightarrow (Z_1^2 + Z_2^2)^{1/2}
   \left( \begin{array}{c} \omega_1 \\ \omega_2 \end{array} \right) 
   (\omega_1, \omega_2).
\label{mhatherm}
\end{equation}
In this form, the mass spectrum can easily be read.  There is a zero 
mode: 
\begin{equation}
\left( \begin{array}{c} - \omega_2 \\ \omega_1 \end{array} \right).
\label{zeromode}
\end{equation}   
plus a massive eigenstate:
\begin{equation}
\left( \begin{array}{c} \omega_1 \\ \omega_2 \end{array} \right) 
\label{massivemode}
\end{equation}
with eigenvalue proportional to $\sqrt{Z_1^2 + Z_2^2}$.

These observations on $\hat{m}$ in (\ref{mhat}) apply at every scale,
but the value of $(\omega_1, \omega_2)$ may differ at different scales.  
The reason is that the basis vectors $\bdelta_1$ and $\bdelta_2$ are 
only required in their definition to be orthogonal to $\balpha$.  Then, 
given the the symmetry $\widetilde{su}(2)_H$, there is in general an 
arbitrariness in choice of which two orthogonal vectors in the plane 
orthogonal to $\balpha$ to be taken as $\bdelta_1$ and $\bdelta_2$.  
Having made a choice at one scale does not oblige one to make the 
same choice at some other scale.  Hence, $(\omega_1, \omega_2)$, the 
components of $\bomega$ in the directions of $\bdelta_1$ and $\bdelta_2$ 
can depend in general on scale, or {\bf [P2]}, as claimed.  

Whether it does change with scale, and if so how, depend first on what 
is taken as $\bomega$, namely on the physical meaning assigned to the 
direction in which the $\widetilde{su}(2)_H$ symmetry is explicitly 
broken.  Secondly, its dependence on scale, if any, may be constrained 
by renormalization group equations in which $\bomega$ figures.  These 
are questions needing closer study to which we have at present no clear 
answer.  

Let us first suppose, however, that $(\omega_1, \omega_2)$ does indeed 
depend on scale or, in the language adopted above, that the direction 
$\bnu' = (- \omega_2, \omega_1)$ in which the zero mode appears does 
rotate with changing scales and ask what will then result.  We are then 
in a situation very similar to that encountered for quarks and leptons 
in the standard sector which we thought we knew how to handle.  Only at 
the end shall we return to ascertain how the result will change when the 
rotation assumption is dropped.

The above comments, from (\ref{calAYQold}) onwards, on the Yukawa coupling 
of co-quarks apply similarly to the Yukawa couplings of the co-leptons but 
these will not be needed in detail in what follows.

\section{Extending CP considerations to include the hidden sector}

Before we attempt to extend CP considerations to the whole of FSM, let us 
first systematize what was done in the standard sector, specifically with 
the view of extending it to include the hidden sector.  The starting point 
there, we recall, was the strong CP problem, or rather the solution of it, 
from which the other results on the CP-violating phase in the CKM and PMNS 
matrices follow.  

We start by noting the following points:

\begin{enumerate}
\item By a solution of the strong CP problem in the FSM, one means a 
rather different process than, say, in the Peccei-Quinn \cite{PecceiQ} 
approach.  In the latter, one starts with the SM with all fermions taken 
massive so that CP is defined already for all fermion states.  Adding 
then to the Lagrangian density a theta-angle term: 
\begin{equation}
-\frac{1}{16 \pi^2} \theta_I \Tr (H^{\mu \nu} H^*_{\mu \nu})
\label{thetatermC}
\end{equation}
for the colour field $H_{\mu \nu}$, with $\theta_I$ of order unity, would 
violate CP even in strong interactions, in contradiction to the existing 
experimental bound on, for example, the neutron electric dipole moment 
which would require a value of $|\theta_I| < 10^{-9}$ \cite{edmneutron}, 
so that a correction to the SM is needed.  Thus, to this end, a new 
$U(1)$ dynamics was introduced in \cite{PecceiQ,weinwil}, to explain why 
the theta-angle prefers to be near vanishing.  The FSM, on the other hand, 
though involving also new dynamics appended to the SM in the form of 
framon fields, these were introduced to understand the mass and mixing 
patterns of fermions, not specifically to solve the strong CP problem.  
The solution of the strong CP problem comes as a bonus since no extra new 
dynamics needs to be introduced to achieve that end.  There is already 
inherent in the FSM at every scale a quark state with zero mass eigenvalue 
for which CP is not yet defined.  A solution of the strong CP problem in 
the FSM means then only the recognition that for an appropriate choice of 
phases in the definition of CP for that quark zero mode would automatically 
induce a chiral transformation to generate a factor in the measure of 
Feynman path integrals \cite{Fugikawa} to cancel the theta-angle term 
that one has started with in the action, leaving thus strong interactions 
CP-invariant.  

\item By strong interactions in the standard sector we mean usually the 
soft interactions between hadrons resulting from the colour interactions 
between quarks.  To isolate the latter, one ignores electroweak effects, 
that is, not only those interactions between quarks resulting from flavour 
and electromagnetism but also all those particles such as leptons which 
have no colour but only electroweak interactions.  Thus to ascertain that 
strong interactions in the standard sector is CP-invariant, it suffices 
to show that QCD for quarks is CP-invariant.

\item Flavour-induced interactions in the standard sector, apart from 
electromagnetism, are termed weak not so much because of the smallness of 
the flavour gauge coupling $g_2$ (which, after all, is only some $\sim 2$ 
times smaller than the colour gauge coupling $g_3$), but because they are 
mediated by massive particles such as $W, Z$ and Higgs, resulting in a 
suppression of their effects by order $g_2^2/M^2$.  It is partly because of 
this, and partly because hadron interactions are soft, that the hard (that 
is, perturbative) flavour-induced interactions of known particles are often 
negligible in hadron physics.  For instance, flavour interactions between 
quarks, being CP-violating via the Kobayashi-Maskawa phase in the CKM 
matrix, do give contributions to the electric dipole moment of the neutron, 
but these are small and still much below the existing, already stringent, 
experimental bound.

\end{enumerate}

With these remarks in mind, let us now formalize the FSM solution of the 
strong CP problem in the standard sector as follows, a procedure which we 
shall then extend to the whole of the FSM, including its hidden sector.  
We first identify and exhibit the approximate theory pertaining to strong 
interactions which conserves CP, and then show how CP is violated as that 
approximation is removed.  

The part of the FSM action involving only those fields appearing in the 
standard sector may be represented schematically as:
\begin{eqnarray}
{\cal A}_{\rm SS} & \sim & FF + GG
+ HH + \theta'_I GG^* + \theta_I HH^* + \nonumber \\
               &&       (D \Phi)^\dagger D \Phi + V[\Phi] 
            + \bar{\psi}_q D \psi_q + \bar{\psi}_\ell D \psi_\ell
            + \bar{\psi}_q \Phi \psi_q + \bar{\psi}_\ell \Phi \psi_\ell,
\label{calASS}
\end{eqnarray}
where $F^{\mu\nu}, G^{\mu\nu}, H^{\mu\nu}$ are respectively the 
$u(1), su(2), su(3)$ gauge fields.  This part of the FSM action differs 
from the SM action only in replacing the Higgs scalar field in SM by 
the flavour framon field {\bf [FF]} where $\Phi$ is a $2 \times 2$ 
matrix, but of this matrix only one column is independent \cite{tfsm}, 
and this column differs from the standard scalar Higgs field $\phi$ in 
SM only by a global vector $\balpha$ as seen in (\ref{PhiF}).  The 
extra factor $\balpha$ does not figure in the potential $V[\Phi]$, which 
is still just the Mexican hat potential for $\phi$ familiar already in 
SM.  As in SM then, the fact that $\phi$ has nonzero vacuum expectation 
value in $V[\Phi]$ gives mass to the Higgs boson $h_W$, and via the term 
$(D \Phi)^\dagger D \Phi$ also to the vector bosons $W$-$Z$.   It also 
gives quarks and leptons a mass matrix of the form (\ref{mfact}), which 
at tree-level has nonzero mass eigenvalue only for the top generation, 
and no up-down mixing.  Nonzero masses for lower generations and mixing 
appear only as higher order effects when $\balpha$ rotates.  All these 
features of the FSM are detailed in earlier publications, for example 
\cite{tfsm}, and are briefly summarized above only for completeness.

Our present focus is on the theta-angle terms coming from instantons 
respectively in flavour $\theta'_I \Tr (G^{\mu\nu}G^*_{\mu\nu})$, and in 
colour $\theta_I \Tr (H^{\mu\nu}H^*_{\mu\nu})$, terms which can lead to 
potentially huge CP-violations.  Suppose for the moment that we are 
interested only in strong effects from quarks, such as the electric 
dipole moment of the neutron, these would be given in terms of Feynman 
path integrals of the form:
\begin{equation}
\frac{\int \delta \psi_q {\cal F}[\psi_q] \exp i {\cal A}_{\rm SS}}
     {\int \delta \psi_q \exp i {\cal A}_{\rm SS}},
\label{FPIofcalF}
\end{equation}
where ${\cal F}[\psi_q]$ is some appropriate functional of the quark 
fields $\psi_q$.  If then, to isolate and exhibit the part of the action 
pertaining only to strong interactions, we can ignore electromagnetic
interactions because of the small coupling and weak flavour interactions 
as per the remark 3 above because of suppression by the mass of the 
flavour bosons.  What we have left in (\ref{calASS}) is only:
\begin{equation}
{\cal A}_{\rm SSstr} \sim H H + \theta_I H H^* + \bar{\psi}_q D_C \psi_q 
                      + \bar{\psi}_q m_q \balpha \balpha^\dagger \psi_q
\label{calASSstr}
\end{equation}
where $D_C$ denotes the covariant derivative with respect to colour only, 
and the last term is the mass matrix in (\ref{mfact}) specialized now 
only to quarks which is obtained by expanding the Higgs field in the 
Yukawa coupling about its vacuum expectation value, keeping only its 
vacuum value and neglecting all higher terms mediated by exchanges of 
the massive Higgs, which would be suppressed.  We note in particular that 
the instanton term $ \Tr(G^{\mu\nu} G^*_{\mu\nu})$ for flavour in the 
numerator of (\ref{FPIofcalF}), like many other terms in the action 
(\ref{calASS}), being uncoupled to the $\psi_q$ in ${\cal F}$ in the 
approximation when flavour interactions are ignored, has been cancelled 
with the same term in the denominator of ({\ref{FPIofcalF}).

The remaining quantity (\ref{calASSstr}), according to (2) above is what 
one would call the strong action in the standard sector, and this 
contains still the instanton term $\Tr (H^{\mu \nu} H^*_{\mu \nu})$ which 
can violate CP hugely.  However, there being zero modes in the mass matrix 
for quarks in (\ref{calASSstr}), an appropriate chiral transformation on 
which will cancel the term $\Tr (H^{\mu \nu} H^*_{\mu \nu})$ in the 
Feynman path integral (\ref{FPIofcalF}), and this is what we called above 
the solution of the strong CP problem in the standard sector in FSM.

Let us turn now to tackle the same problem when extended to include the 
hidden sector.  To do so, we shall need to add to (\ref{calASS}) terms 
involving the new fields which had not figured in the standard sector 
before, thus (again schematically):
\begin{eqnarray}
{\cal A} & \sim & {\cal A}_{\rm SS} + \bV[\bPhi] + \nu[\Phi, \bPhi] +
              (D \bPhi)^\dagger D \bPhi \nonumber \\
         &   & + \bar{\psi}_Q D \psi_Q + \bar{\psi}_L D \psi_L 
               + \bar{\psi}_Q \bPhi \psi_Q + \bar{\psi}_L \bPhi \psi_L.
\label{calA}
\end{eqnarray}
In particular, the framon self-interaction potential constructed from the 
underlying invariance of FSM becomes much more intricate, involving now 
the colour framon in (\ref{PhiC}) which is a full $3 \times 3$ matrix with 
many more field degrees of freedom.  Details of this can be found in, for 
example, \cite{tfsm}.  Besides the Mexican hat potential $V[\Phi]$ for the 
standard Higgs field included already in (\ref{calASS}), we have now also 
a $\bV$ for $\bPhi$, as well as a term $\nu[\Phi, \bPhi]$ which links the 
flavour and colour framon fields.  As a result, the vacuum also becomes 
much more intricate, the details for which are given also in \cite{tfsm}.  
However, what matters here is just the fact that the colour framon has a 
nonzero vacuum expectation value, giving thus masses to the scalar bosons 
H, and also via the term $(D \bPhi)^{\dagger} D \bPhi$ masses to the 
vector bosons G.  These masses, as suggested by the analysis in 
\cite{cfsm}, are typically of order TeV.  Hence, according to point 3 
above, for co-quarks and co-leptons, it is now the colour interactions via 
the exchange of these massive Hs and Gs which will be suppressed and are 
to be considered ``weak'', in the same spirit as the flavour interactions 
of quarks and leptons in the standard sector were considered weak.  But 
it is the flavour interactions, coming from the local flavour that 
co-quarks still carry, which are to be taken as ``strong'' instead.  
Local flavour in the hidden sector, being unbroken and confining, can 
thus form, from co-quarks (and anti-co-quarks) via flavour confinement, 
co-hadrons, which can have, among themselves, soft interactions similar 
in strength to the soft interactions we see among hadrons in the standard 
sector \cite{fsmpop,cfsm}.  In other words, as anticipated, the roles of 
colour and flavour are switched from the standard to the hidden sector.

In parallel then to (\ref{FPIofcalF}), strong effects from co-quarks 
would be given in terms of Feynman path integrals of the form:
\begin{equation}
\frac{\int \delta \psi_Q {\cal F}'[\psi_Q] \exp i {\cal A}}
     {\int \delta \psi_Q \exp i {\cal A}},
\label{FPIofcalF'}
\end{equation}
where ${\cal F}'[\psi_Q]$ is some appropriate functional, now of the 
co-quark field $\psi_Q$.  And if, to identify the strong action, as 
was done above for the standard sector, we omit for the moment the 
electromagnetic interactions of the co-quarks $Q$ because of the small 
coupling, and their weak colour interactions because of their suppression 
by the H and G mass, as well as those particles which have only these 
interactions, what we have left of ${\cal A}$ which is relevant for the 
hidden sector in the path integral (\ref{FPIofcalF'}) is just:
\begin{equation}
{\cal A}_{\rm HSstr} \sim G G + \theta'_I G G^* + \bar{\psi}_Q D_F \psi_Q 
     + \bar{\psi}_Q m_Q \psi_Q,
\label{calAHSstr}
\end{equation} 
where $m_Q$ is the mass matrix of the co-quark $Q$:
\begin{equation}
m_Q = \zeta_S \left( \begin{array}{cc} \third (1 - R) \hat{m} & 0 \\
                0 & \twothirds (1 + 2R) \end{array} \right),
\label{mQ}
\end{equation}
with $\hat{m}$ as given in (\ref{mhatherm}) of the preceding section, $D_F$ 
the covariant derivative with respect to flavour only, and $\zeta_S$ and 
$R$ are scalar quantities given in \cite{tfsm}, for example, which, though 
important for consideration of the actual mass spectrum and so on, are not 
necessary for the discussion on CP here.  There are terms in ${\cal A}$ 
other than those exhibited in (\ref{calAHSstr}), but they are either 
negligible in the present approximation or else, being linked to $Q$ only 
by the neglected interactions, would cancel in (\ref{FPIofcalF'}) between 
the numerator and the denominator.

There is still in (\ref{calAHSstr}) the term $\theta'_I \Tr (G^{\mu \nu} 
G^*_{\mu \nu})$, which can potentially give large CP-violating effects to 
strong interactions of the co-quarks $Q$, as represented by (\ref{FPIofcalF'}).  
However, given that the mass matrix $m_Q$, as shown in the analysis around 
(\ref{mhatherm}) in the last section, has a mode, say $Q_0$, in the direction 
$\bnu'$ which has zero mass eigenvalue, a chiral transformation can be 
performed on that mode, thus:
\begin{equation}
\psi_{Q_0} \rightarrow \exp (-i \half \theta'_I \gamma_5) \psi_{Q_0}
\label{chitransQ0}
\end{equation}
to generate a factor from the measure of the integral (\ref{FPIofcalF'}) to 
cancel the term $\theta'_I \Tr (G^{\mu \nu} G^*_{\mu \nu})$ from (\ref{calAHSstr}),  
so as to leave the effects of strong (flavour) interaction of co-quarks 
CP-invariant.

At first sight, this result may not seem too surprising, given that the 
standard and hidden sectors in the FSM are related by having the roles 
of colour and flavour interchanged, and both quarks and co-quarks are 
known to have zero modes.  Hence, if in the standard sector, the strong 
CP problem can be cured by a chiral transformation on the quark zero 
mode, one may not perhaps be too surprised to find that in the hidden 
sector, the parallel strong CP problem can be cured by a parallel chiral 
transformation on the co-quark zero mode.  On closer look, however, one 
finds that the situation is in fact more intricate.  For solving the 
strong CP problem in the standard sector, the chiral transformation is 
performed on the quark zero mode in a direction $\bnu$ in generation 
or dual colour space.  One would expect then that, in parallel under a 
colour-flavour interchange, the strong CP problem in the hidden sector 
would be cured by a chiral transformation on a co-quark zero mode in 
some direction in ``co-generation'' or dual flavour space.  However,
according to the analysis in \cite{cfsm}, no ``co-generation'' exists in 
the hidden sector parallel to generation in the standard sector, because 
of a so-called minimal embedding condition imposed on the flavour framon 
{\bf [FF]} which was what led to one column in $\Phi$ in (\ref{PhiF}), as 
explained there, being made redundant.  Instead, in the above treatment, 
the direction $\bnu'$, containing the zero mode $Q_0$ on which the chiral 
transformation is performed to cure the strong CP problem in the hidden 
sector, is still in dual colour space, or rather in a subspace of that 
space orthogonal to the vector $\balpha$.  Nevertheless, despite this 
difference in details, it seems that the parallel in result still holds.     

For the FSM as a whole then, including both the standard and the hidden 
sectors, strong interactions can be made CP-invariant at tree-level (in 
fact even at one-loop level at any fixed scale) by appropriately defining 
CP for the zero modes of the quarks and co-quarks.  This was done in the 
approximation when electromagnetic and weak interactions (that is, flavour 
in ${\bf [SS]}$ and colour in ${\bf [HS]}$) were omitted.  

Let us next put back these interactions so far ignored to consider CP 
for the FSM in full.  Quarks having then recovered their weak flavour 
interactions, the chiral transformation (\ref{thetatermC}) performed on 
the quark state in the $\bnu$ direction to cure the strong CP-problem in 
${\bf [SS]}$ will then generate in the measure of the full Feynman path 
integrals a factor with exponent:
\begin{equation}
- \frac{\theta'_C}{16\pi^2} \Tr (G^{\mu \nu} G^*_{\mu \nu}),
\label{thetatermFC}
\end{equation}
with $\theta'_C = - 3 \theta_I/4$, as noted in \cite{cpslept}.  This 
can lead to large, experimentally unwanted, CP-violations in electroweak 
effects.  However, as was shown in \cite{cpslept}, (\ref{thetatermFC}) 
can again be cancelled by another factor generated in the measure of full 
Feynman path integrals by a chiral transformation on the lepton state in 
the $\bnu$ direction, making then the standard sector, including now all 
interactions, fully CP-invariant at any fixed scale.

Similarly, in the hidden sector, the co-quarks, having recovered their 
weak colour interactions, would generate, via their earlier chiral 
transformation to cure the strong CP problem in ${\bf [HS]}$, from the 
measure of full Feynman path integrals a factor with exponent:
\begin{equation}
- \frac{\theta_F}{16\pi^2} \Tr (H^{\mu \nu} H^*_{\mu \nu}),
\label{thetatermCF}
\end{equation}
with $\theta_F = - \theta'_I/2$.  Such a term could in turn give large 
CP-violations to the weak colour interactions in the hidden sector but 
can again be cancelled by making the appropriate chiral transformations 
on the zero mode(s) of the co-leptons.  However, our knowledge on the 
spectrum of co-leptons being so meagre, no attempt at further elucidation 
is at present possible.

The arguments above being quite involved and loaded by necessity with 
details, it is worthwhile for clarity to go succinctly over them once 
more, even at the cost of some repetitions, as follows. 

Schematically, the strong part of the FSM action appears as:
$$
{\cal A}_{\rm str}={\cal A}_{\rm SSstr}+{\cal A}_{\rm HSstr},
$$
where
\begin{eqnarray*}
{\cal A}_{\rm SSstr} & \sim & HH + \theta_I HH^* + \bar{\psi}_q \not{
\hspace*{-1mm}D}_{\rm C} 
\psi_q + \bar{\psi}_q m_q \balpha \balpha^\dagger \psi_q, \\
{\cal A}_{\rm HSstr} & \sim &GG + \theta'_I GG^* + \bar{\psi}_Q 
\not{\hspace*{-1mm}D}_{\rm F} 
\psi_Q + \bar{\psi}_Q m_Q  \psi_Q. 
\end{eqnarray*}

There are two CP violating terms: $\theta_I HH^*$ involving colour for the SS and 
$\theta'_I GG^*$ involving flavour in the HS.  They could produce large CP violation
in respectively the SS and the HS.
\begin{itemize}
\item[[$1\!\!\!$]] The $\theta_I HH^*$ term can be cancelled in Feynman integrals 
by a chiral transformation $\alpha$ on the 2 zero modes (1 up, 1 down) of the 
quark mass matrix, with
$$
  \alpha = -\quarter \theta_I.
$$
\item[[$2\!\!\!$]] Assuming that large CP violation in the HS is not acceptable,
the $\theta'_I GG^*$ term can be cancelled by a chiral transformation 
$\alpha'_{\rm HS}$ on the zero mode of the co-quark mass matrix, with
$$
  \alpha'_{\rm HS}=-\half \theta'_I. 
$$
\end{itemize}
These two steps will make the strong action manifestly CP invariant.  

Next we put back the weak parts of the action, but now with two additional terms 
generated by the chiral transformations [1] and [2] above.
\begin{itemize}
\item[[$3\!\!\!$]] The chiral transformation $\alpha$ on the quarks, the LH 
components only of which carry flavour, will generate from the measure of 
Feynman integrals an extra term
$$
 - \frac{\theta'_C}{16\pi^2} \Tr GG^* = 
-\frac{3 \alpha}{16\pi^2} \Tr GG^* = \frac{\threequarters \theta_I}{16\pi^2} 
\Tr GG^*,
$$
with a factor of 3 because quarks come in 3 colours, as explained in
\cite{cpslept}.  This extra term can be cancelled in the Feynman integrals 
by a chiral transformation $\alpha'$ on the zero modes of the lepton mass 
matrix, with $$ \alpha'=-\theta'_C=\threequarters \theta_I.
$$
\item[[$4\!\!\!$]] Similarly a chiral transformation $\alpha'_{\rm HS}$ on the 
co-quarks will generate a term $-\frac{\theta_F}{16\pi^2} \Tr HH^*$ 
in the action, with
$$
\theta_F =  2 \alpha'_{\rm HS} = - \theta'_I, 
$$
with a factor of 2 because co-quarks come in 2 flavours.  This additional 
CP violating term can also in principle be cancelled by a chiral 
transformation $\alpha_{\rm HS}= - \theta_F = \theta'_I$ on the co-leptons, 
but this we leave open at present, lacking sufficient knowledge of the 
co-lepton spectrum. 
\end{itemize}

Hence, we see that by suitable chiral transformations on the zero modes of the
relevant mass matrix, all CP violating terms have been cancelled, which means
that CP is conserved at tree-level, both in the SS and the HS.  However, when 
loop corrections are included, the zero modes, on which the above chiral 
transformations act, will change with scale and induce CP-violating phases in the 
CKM and the PMNS matrices.  Both these effects are observed in experiments, the 
first well-established and the second a 3 $\sigma$ effect more recently.  For 
HS, on the other hand, though parallels can apply, little can be said at this 
stage for lack of expermental knowledge.

There are two points on the above treatment that we wish to highlight for 
later reference:
\begin{itemize} 
\item (I) CP is conserved at every scale but violations set in when scale 
changes are involved;
\item (II) These CP-violations come from radiative corrections and are 
thus bound to be perturbatively small.
\end{itemize}

These results have been deduced under the assumption that $\bomega$ in 
the Yukawa coupling (\ref{calAYQnew}) rotates with scale.  What will be 
changed if this assumption is dropped?  Now it is clear that so long as 
the zero mode in the hidden sector exists, given (\ref{calAYQnew}), 
then the theta-angle term $\theta'_I \Tr (G^{\mu \nu} G^*_{\mu \nu})$ in 
(\ref{calAHSstr}) can be cancelled by a chiral transformation on that 
zero mode to keep strong interactions in the hidden sector CP invariant, 
independently of whether the state vector of that zero mode rotates with 
scale or not.  With no rotation, however, one can no longer appeal to 
the ``leakage mechanism'', used in for example \cite{tfsm} for quarks and 
leptons, to give nonzero physical masses or residual CP-violations in 
mixing matrices to co-quarks and co-leptons.  Massless particles may then 
be unavoidable in the hidden sector, which last will then be CP-invariant 
at all scales.  These changes would affect the physical consequences to 
be drawn in the next section, as we shall indicate at the end of that 
section.

\section{Remarks}

\subsection{The CP-violating phase $\delta'_{CP}$ in the PMNS matrix}

An immediate consequence of the above extension of CP to the hidden sector 
is to turn the following result of \cite{cpslept} from a part-estimate 
into a prediction of the actual value for the Jarlskog invariant of the 
PMNS matrix,
\begin{equation}
J' \sim - 0.012.
\label{J'est}
\end{equation}
We recall that there are actually two theta-angle terms for flavour, namely
\begin{equation}
-\frac{1}{16 \pi^2} (\theta'_I + \theta'_C) \Tr (G^{\mu \nu} G^*_{\mu \nu}), 
\label{thetatermsF}
\end{equation}
where the $\theta'_I$ term comes from instantons and appears in the original 
action while $\theta'_C$ comes from the measure in Feynman path integrals as 
the result of a chiral transformation made on the quark zero mode to cure 
the strong CP problem in QCD.  The angle $\theta'_I$ is unknown, but an estimate
for the value of $\theta'_C$ has been obtained from a earlier fit to data 
\cite{tfsm}.  Now, to cancel both these terms so as to make the whole theory 
CP-invariant at a fixed scale, there was available in \cite{cpslept}, where 
the discussion was restricted to the standard sector, only the one chiral 
transformation on the lepton zero mode.  Hence, the Jarlskog invariant which 
ensues for leptons would depend on the yet unkown value of $\theta'_I$, resulting 
thus in only a part-estimate, or at best an order-of-magnitude estimate.  
However, now that the discussion has been extended to the hidden sector, one 
finds that in order to guarantee that there should be no strong CP problem 
there, the $\theta'_I$ term has to be cancelled by a chiral transformation 
on the co-quark zero mode.  This leaves only the known term $\theta'_C$ to 
be cancelled by the chiral transformation on the lepton zero mode, giving 
then the result (\ref{J'est}) as a prediction for the actual value of the 
Jarlskog invariant in the PMNS matrix for leptons. 

And as shown already in \cite{cpslept}, (\ref{J'est}) is (i) has the right 
order of magnitude, (ii) has the right sign and (iii) has a value in the 
range favoured by present experiment, providing thus, it would seem, a 
nontrivial check on the FSM scenario above, somewhat fanciful though the 
latter might perhaps appear. 

We should recall also that the turning of (\ref{J'est}) into a prediction 
for the value of $J'$ depends on the condition that strong interactions 
in the hidden sector is CP-conserving.  For this, of course, one has no 
direct evidence, since even the existence of the hidden sector has yet to 
be ascertained.  One can cite, however, the following points in support:
\begin{itemize}
\item (a) One has, of course, first to define what CP is before one can 
speak of CP-violations.  It seems both logically and aesthetically unsound 
thus not to make a theory CP-conserving at the strong interaction level 
where a simple choice of phase is enough to achieve that end.  And one is 
free to do so in the definition of CP for certain zero modes in the FSM, 
as was done above.
\item (b) Given that the bound on CP-violations in strong interactions in 
the standard sector, as provided by the experimental bound on the electric 
dipole moment of the neutron \cite{KGreen}, is extremely stringent, and 
that the hidden sector is not entirely cut off from the standard one but 
is connected to it by some portal states, as shown in \cite{cfsm}, strong 
CP-violations in the hidden sector, if allowed, would leak into the 
standard sector as well, which is likely to cause problems in satisfying 
the bounds set by the neutron dipole moment, although one is not in a 
position as yet to verify that this is indeed the case.  
\item (c) CP-violations at strong interaction level in the hidden sector 
might lead to a superabundance of dark matter in the universe much beyond 
what is needed.  See note (c) of the subsection below.

\end{itemize}

\subsection{Hints on the material content of the universe}

Two outstanding and quite novel features [I] and [II] of CP in the FSM 
are listed near the end of the section above. 

We recall now that in understanding the material content of the universe, 
CP-violations come in  as a crucial factor, as pointed out by Sakharov 
already in 1967 \cite{Sakharov}.  Roughly, the picture one has is the 
following.  At very early times when the universe was very hot, masses 
were negligible and massive particles were present as abundantly as 
massless ones.  When the universe cooled, however, massive particles like 
baryons started to annihilate with their C-conjugates, eventually almost 
all into photons, leaving only a small excess of baryons over anti-baryons, 
and this is most of the luminous matter that we now see.  For this excess 
of baryons over anti-baryons to occur, CP had, according to Sakharov, to be 
violated at some stage.  That CP-violations are weak effects then explain 
why this excess of baryons over anti-baryons is small and why photons occur 
as much as a billion times in numbers as baryons do in the universe today. 

However, now that the FSM has injected the above two novel features into 
CP-violations, the above picture for the material content of the univese 
is also given some new twists.
\begin{itemize}
\item (a) In the FSM, (I) says that although the theory can be made at 
any scale to be CP-conserving, CP-violations would automatically develop 
as the scale changed (as the universe cooled), so that there seems to be 
no need to assume any pre-existing (or primordial) CP-violation in the 
universe for the picture outlined above to materialize.
\item (b) Most models for baryo- and lepto-genesis would take as input 
CP-violations from the measured CKM matrix (and presumably also the PMNS 
matrix now that it is beginning to be seen), but in the FSM the CKM and 
PMNS matrices are supposed to be themselves the consquence of (I) above.  
It would seem therefore, that a study of the effect directly from (I) on 
baryo- and lepto-genesis might give some new insight.
\item (c) Parallel considerations to those above for the standard sector 
would suggest by (I) that massive particles in the hidden sector would 
also annihilate into photons when the universe cooled, leaving as residue 
by (II) only a small excess of particles over anti-particles.  In other
words, given that most hidden sector particles are expected to end up as 
dark matter, this would seem to mean that the overwhelming predominance 
in number of photons over massive matter particles would be maintained in 
the dark sector as it was in the luminous sector.  But this might not be 
the case without (II), that is, if we had allowed CP to be violated at 
the strong interaction level in the hidden sector.
\item (d) Given that the mechanisms suggested above for CP-violation 
are similar in the two sectors, it is tempting to assume that roughly 
the same amount of matter would survive annihilation when the universe 
cooled.  Recall now that colour framons {\bf [CF]} in (\ref{PhiC}) number 
more in degrees of freedom than flavour framons {\bf [FF]} in (\ref{PhiF}) 
(of which, as noted there, only one column needs be kept \cite{tfsm}) by 
a ratio of 9:2; and that colour framons are what make up matter in the 
hidden sector as opposed to flavour framons which make up matter in our 
standard sector.  Then, if we really assume that the same proportion in 
each sector would survive annihilation when the universe cooled, we would
end up with the same ratio $9/2 = 4.5$ of dark matter in the hidden sector 
to luminous matter in our standard sector (i.e.\ dark matter making up 82 
percent of the total).  This is not far from what is given in \cite{pdg} 
for $\Omega_{cdm}/\Omega_b \sim 5.3$ (or dark matter making up 84 percent 
of the total).  The assumptions one started with, however, are much too 
simplistic for this coincidence, though interesting, to be taken seriously
at present.  First, CP-violation, though necessary, is not the only factor 
that would govern the amount of excess matter that survived annihilation. 
And secondly, in the hidden sector, one does not even know the particle 
spectrum well enough to decide which particles are likely to dominate as 
dark matter.  Even if one assumes that, in parallel to the standard sector 
where baryons dominate, co-baryons also dominate as dark matter in the 
hidden sector, sufficient differences in property exist between baryons 
and co-baryons to dissuade one from drawing too close a parallel between 
them.  \footnote{For instance, co-baryons are binary objects, each formed 
from 2 co-quarks by flavour $su(2)$ confinement while baryons are trinary, 
each formed from 3 quarks by colour $su(3)$ confinement.  This means that, 
if the formation of either had to compete with other processes, co-baryons 
could end up in larger numbers than baryons would.  And besides, co-baryons 
are bosonic while baryons are fermionic, adding thus further food for 
thought.}   
\end{itemize}

\vspace{.5cm}

Again, we need to ask what would change if the assumption of rotation for 
the zero mode $\bnu'$ in the hidden sector is dropped.  In view of the 
observations made in the last paragraph in the preceding section, one sees
that everything in this section is left unchanaged except for (c) and (d) 
of subsection 4.2.   Without this assumption, one will need another source 
for CP-violation in the hidden sector for massive particles to survive 
annihilation when the universe cooled.  Moreover, co-quarks and co-leptons 
with vanishing physical masses may no longer be avoidable.  Now although 
one has no direct evidence against zero mass co-quarks or co-leptons, 
their existence may give more hot dark matter in the universe than one 
would like.

\end{document}